\documentclass[aps,amsmath,prl,superscriptaddress,a4paper,floatfix,showpacs,twocolumn]{revtex4-1}
\usepackage{graphicx}
\usepackage{amsmath}
\usepackage{amsfonts}
\usepackage{amssymb}

\newcommand{\beq}{\begin{equation}}
\newcommand{\eneq}{\end{equation}}
\newcommand{\bea}{\begin{eqnarray}}
\newcommand{\enea}{\end{eqnarray}}

\begin{document}
\title{Direct transition from quantum escape to phase diffusion regime in YBaCuO biepitaxial Josephson Junctions}

\author{Luigi Longobardi}
\email{llongobardi@ms.cc.sunysb.edu}
\affiliation{Seconda Universit\'{a} degli Studi di Napoli, Dipartimento di Ingegneria dell'Informazione, via Roma 29, 81031 Aversa (Ce) Italy}
\affiliation{CNR-SPIN UOS Napoli, Complesso Universitario di Monte Sant'Angelo via Cinthia, 80126 Napoli (Na) Italy}
\author{Davide Massarotti}
\affiliation{Universit\'{a} degli Studi di Napoli "Federico II", Dipartimento di Scienze Fisiche, via Cinthia, 80126 Napoli (Na) Italy}
\affiliation{CNR-SPIN UOS Napoli, Complesso Universitario di Monte Sant'Angelo via Cinthia, 80126 Napoli (Na) Italy}
\author{Daniela Stornaiuolo}
\affiliation{CNR-SPIN UOS Napoli, Complesso Universitario di Monte Sant'Angelo via Cinthia, 80126 Napoli (Na) Italy}
\author{Luca Galletti}
\affiliation{Universit\'{a} degli Studi di Napoli "Federico II", Dipartimento di Scienze Fisiche, via Cinthia, 80126 Napoli (Na) Italy}
\affiliation{CNR-SPIN UOS Napoli, Complesso Universitario di Monte Sant'Angelo via Cinthia, 80126 Napoli (Na) Italy}
\author{Giacomo Rotoli}
\affiliation{Seconda Universit\'{a} degli Studi di Napoli, Dipartimento di Ingegneria dell'Informazione, via Roma 29, 81031 Aversa (Ce) Italy}
\author{Floriana Lombardi}
\affiliation{Dept of Microtechnology and Nanoscience, Chalmers University of Technology, S-41296 G\"{o}teborg, Sweden.}
\author{Francesco Tafuri}
\affiliation{Seconda Universit\'{a} degli Studi di Napoli, Dipartimento di Ingegneria dell'Informazione, via Roma 29, 81031 Aversa (Ce) Italy}
\affiliation{CNR-SPIN UOS Napoli, Complesso Universitario di Monte Sant'Angelo via Cinthia, 80126 Napoli (Na) Italy}

\date{\today}

\begin{abstract}

Dissipation encodes interaction of a quantum system with the environment and regulates the activation regimes of a Brownian particle. We have engineered grain boundary biepitaxial YBaCuO junctions to drive a direct transition from quantum activated running state to phase diffusion regime. The cross-over to the quantum regime is tuned by the magnetic field and dissipation is encoded in a fully consistent set of junction parameters. To unravel phase dynamics in moderately damped systems is of general interest for advances in the comprehension of  retrapping phenomena and in view of quantum hybrid technology.

\end{abstract}


\maketitle

The impressive development of superconductive systems in the field of quantum information processing and the expertise gained on manipulating coherent entangled states and different coupling regimes with the environment \cite{clarke2008} have boosted research on several complementary aspects of coherence and dissipation.
Due to their design scalability and to the flexibility in controlling the level of damping, Josephson systems have proven to be a fantastic test-bench for studying fundamental physics problems such as the quantum superposition of alive and dead states of a Schr\"{o}dinger's cat\cite{friedman}, the behavior of an artificial atom in cavity quantum electrodynamics experiments\cite{wallraff2004}, or measurements of quantum coherence in macroscopic systems\cite{nakamura,chiorescu}.

Both the time evolution of the position of a Brownian particle and the electrodynamics of a Josephson junction (JJ)\cite{barone,likharevbook} can be described by a Langevin equation \cite{langevin} of the form
\begin {equation}
\ddot{\varphi}+\dot{\varphi}/Q+ dU/d\varphi = \xi(t)
\label{lang}
\end {equation}
In this equation the time is normalized to $1/\omega_p$, with $\omega_p=(2eI_{co}/\hbar C)^{1/2}$ representing the plasma frequency at zero bias current and $I_{co}$ and C being the junction critical current in absence of thermal fluctuations and capacitance, respectively. The potential $U$ is the well known periodic "washboard" potential associated to the dynamic of a JJ, $U(\varphi)=-E_J\left(\cos\varphi +\frac{I}{I_{co}}\varphi\right)$, where $\varphi(t)$ is the superconductive phase and $E_J=\hbar I_{co} /2e$ is the Josephson energy.  $\xi(t)$ is a white noise driving force such that
\begin{equation}
\left<\xi(t)\right>=0; \hspace{5 mm} \left<\xi(t),\xi(t')\right>=\sqrt{k_B T/Q E_J}\delta(t-t').
\end{equation}

\begin{figure*}[htbp]
\begin{center}
\includegraphics[width=\linewidth]{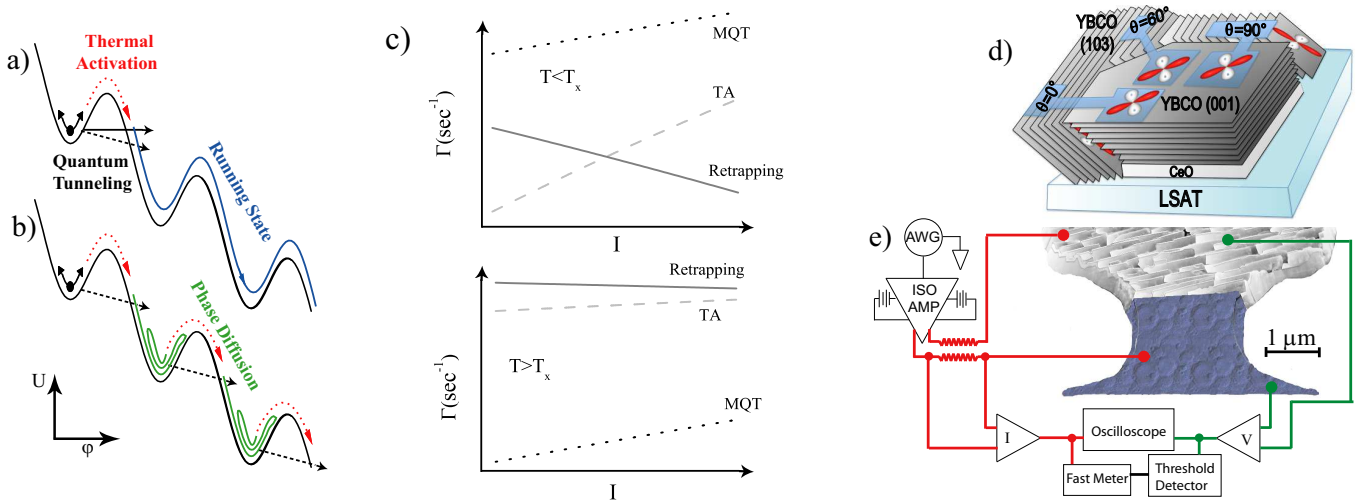}
\caption{ a)  Thermal (red dashed line) or quantum activated escape in the tilted periodic potential. Quantum escape is represented both for very low ideal ($Q >> 1$ - continuous black line) and high ($ 1< Q <5$ - dashed black line) levels of dissipation respectively.  b) Diffusive motion due to multiple escapes and retrapping in subsequent potential wells. c) Schematic escape rates of MQT, TA and retrapping processes at $T<T_x$ (top panel) and $T>T_x$ (bottom panel) d) Sketch of the off-axis biepitaxial junctions.  The JJ is formed at the boundary between (001) YBaCuO and (103) YBaCuO electrodes. Three interface orientations $\theta = 0^o$, $60^o$  and $90^o$ are shown as examples. Different interface orientations (lobe vs. lobe, node vs. lobe and any configuration in between) can be achieved with proper patterning of the seed layer \cite{nuovo}. e)Picture of the device along with a block diagram of the experimental setup. In the bottom part of the image (in blue) is the YBaCuO (001) electrode while in the top part (gray) the needle-like YBaCuO (103) grains are visible.}
\label{potential}
\end{center}
\end{figure*}

The parameter $Q$ is given by $Q=\omega_p R C$ \cite{barone,likharevbook} with R the shunt resistance. In a more general approach $Q$ has a frequency dependence\cite{kautz90} better responding to the need of including external shunting impedance. In our case the simplest approach based on the Q factor calculated only at plasma frequency\cite{kivioja2005,mannik2005}, gives a satisfying  account for experimental data, as demonstrated below. It is a natural choice to use a JJ in order to study the Brownian motion of a particle in a dissipative tilted periodic potential (see Fig.\ref{potential}a,b).

In this letter we demonstrate a direct transition from a running state, obtained following a quantum activation, to diffusive Brownian motion in YBaCuO JJs. Multiple retrapping processes in subsequent potential wells characterize phase regimes where diffusive phenomena play a relevant role \cite{kautz90,vion,kivioja2005,mannik2005,krasnov2005}(see Fig.\ref{potential}b). The relevant parameters driving the occurrence of these phenomena are the operational temperature $T$, the damping factor $Q$ and the critical current $I_{co}$. The various operation scenarios for a JJ have been condensed in a phase diagram by Kivioja et al.\cite{kivioja2005} who have shown that by spanning the ($E_J, k_BT$) parameter space it is possible to engineer all different regimes ranging from phase diffusion and thermal activation to macroscopic quantum tunneling (MQT). MQT takes place not only for low values of dissipation ($Q >> 1$), but also for intermediate levels of dissipation ($1< Q < 5$). We explore a new region of this phase diagram, made available by the different ranges of $I_{co}$ and of the standard deviation of the switching distribution $\sigma$ offered by these junctions when compared with most low temperature superconductors (LTS) JJs. The moderately damped systems are particularly significant and promising to address and quantify interactions of a quantum system with the environment  \cite{Leggett} which is, apart from the its intrinsic interest for fundamental physics, a cornerstone for the development of whatever quantum hybrid technology.

The experimental observation of the various regimes in a JJ is based on the measurement of the switching current distribution (SCD) and the study of the behavior of its first and second momenta (the mean $\overline{I}$ and the width $\sigma$) as function of temperature. In an underdamped junction ($Q>10$) \cite{devoret1985}, below a crossover temperature $T_x$ the escape process is mostly due to MQT (see Fig. \ref{potential}a), marked by a temperature independent $\sigma$, while above $T_x$ the process of escape is due to thermal activation (TA) above the potential barrier, with a distinctive increase of $\sigma$ with temperature. In moderately damped junctions \cite{kautz90,kivioja2005,mannik2005,krasnov2005,fenton2008,Bae2009,luigi} with $2<Q<5$ a transition from TA to 'phase diffusion' (PD) regime occurs at a crossover temperature $T^* > T_x$. $T^*$ marks a distinctive change in the sign of the temperature derivative of $\sigma$, with  $d\sigma/dT>0$ for $T <T^*$ and $d\sigma/dT<0$ for $T > T^*$. When escape out of a well occurs at too low currents in PD regime\cite{kautz90}, the energy gained by passing from one well to the next one barely exceeds the dissipative losses and the particle eventually gets retrapped, basically diffusing to the very next wells. The SCD histograms move to lower currents I till they touch the limit  $I_R=(4*I_{co}/\pi)1/Q$ in the most common cases with $Q\gg1$\cite{stewart,kautz90}.



\begin{figure*}
\begin{center}
\includegraphics[width=\linewidth]{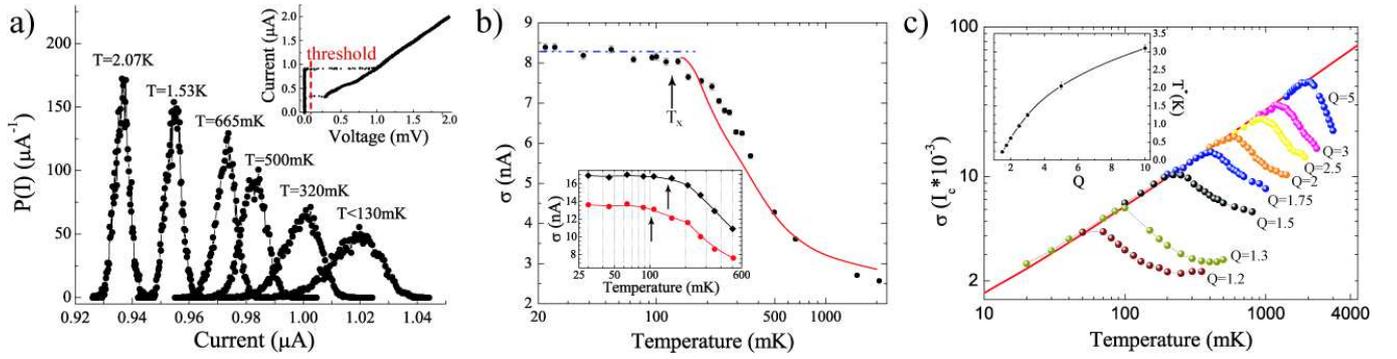}
\caption{a) Measured switching current probability distribution P(I) at different bath temperatures, for sample A. The inset shows the device current voltage characteristic measured at 30mK. The reference value for the threshold detector is also displayed. b) Temperature dependence of the standard deviation, $\sigma$, of the switching distributions for sample A. The dash-dotted line marks the temperature-independent SCD widths in the quantum tunneling regime, the red solid line is the results of simulations in the diffusive regime with a damping parameter of Q=1.3. The inset shows temperature dependent data for sample B acquired at two different values of the applied magnetic field. c) Simulated thermal behavior of the width of the switching histogram for several values of the Q damping parameter.
In the inset we report the dependence of the turn-over temperature $T^*$ on the damping parameter.}
\label{qpd4}
\end{center}
\end{figure*}

We report measurements of SCD in a temperature range from 20mK to 2.2K. Our data are characterized by two distinct regimes. Below 135mK the widths of the SCDs show no significative variation. This is a typical signature of a quantum activation regime. Above 135mK the negative temperature derivative of $\sigma$ is consistent with a diffusive motion due to multiple retrapping in the potential wells. This regime has been fitted using existing theories on phase diffusion\cite{fenton2008} with a  damping factor $Q= 1.3$.
The observations are qualitatively consistent with the escape rates sketched in Fig.\ref{potential}c.
We have substantially engineered a device with  $T^* \lesssim T_x$. For temperatures T well below $T_x$, MQT contributions to escape rates are larger than those coming from both thermal escape and multiple retrapping processes (Fig.\ref{potential}c top panel) differently from the case  $T > T_x$ (Fig.\ref{potential}c bottom panel). The contiguity between quantum escape ($T<T_x$) and phase diffusion ($T>T_x$) leads to MQT phenomena characterized by low Q values and not necessarily to quantum phase diffusion. This MQT process can be represented in Fig. \ref{potential}a as a dashed line to manifest interaction with the environment \cite{Ovch}, and responds to what it can be possibly experimentally ascertained.  This phenomenology is quite distinct from  all previous studies \cite{kautz90,kivioja2005,mannik2005,krasnov2005,fenton2008,Bae2009,luigi}, where in the transition to quantum activation, retrapping processes decay faster than thermal escape, and from the work of Yu et al. \cite{yu_2011}, where  the occurrence of a quantum activated phase diffusion has been claimed. In ref \cite{yu_2011},  the semiclassical nature of their quantum phase diffusion is testified by the dependence of $\sigma$ on the temperature over the entire temperature range, and the transition is as a matter of fact revealed by a change of the temperature derivative of $\sigma$ \cite{yu_2011}. MQT processes are substantially followed and assisted by thermally-ruled retrapping processes. However, a fully quantum account of phase fluctuations passes through the 'empirical' condition of a Josephson energy much larger than Coulomb energy, $E_J >> E_C$ with $E_C=e^2/2C$ (see below),  given by Iansiti et al. \cite{iansiti}.  This condition does not occur both in this experiment  and in the work of Yu et al. \cite{yu_2011}, which both represent complementary significant advances towards the observation of a fully-quantum phase diffusion, better defining its domain through the border with competing processes.

We have used YBaCuO off-axis GB biepitaxial JJs, whose scheme is shown in Fig. \ref{potential}d \cite{nuovo,rop,bauch_2005}. The GB is determined at the boundary between the 103-oriented grains growing on the bare substrate and the 001 grain growing on the $CeO_2$ seed layer. We have engineered junctions on $(La_{0.3}Sr_{0.7})(Al_{0.65}Ta_{0.35})O_3$ (LSAT) rather than $SrTiO_3$ (STO) substrates, where MQT in a high temperature superconductor JJ was first demonstrated\cite{bauch_2005}. The new design fully responds to the need of reducing stray capacitances and of better isolating the GB behavior\cite{daniela}. Specific capacitances are one order of magnitude lower than those measured on STO-based devices\cite{bauch_2005,daniela}. Dynamical junction parameters can be tuned by choosing the interface orientation indicated by the angle $\theta$ in Fig. \ref{potential}d, which also sets d-wave induced effects\cite{nuovo,rop}.

To study the escape rates of YBaCuO Josephson junctions we have thermally anchored the sample to the mixing chamber of a He3/He4 Oxford dilution refrigerator and performed measurements of the junction switching current probability. A schematic representation of our experimental set-up is shown in Fig.\ref{potential}e. A full description of the apparatus is discussed in detail elsewhere \cite{luigi}. Filtering is guaranteed by  a room temperature electromagnetic interference filter stage followed by low pass RC filters with a cut-off frequency of 1.6MHz anchored at 1.5 K, and by a combination of copper powder and twisted pair filters thermally anchored at the mixing chamber of the dilution refrigerator. The bias current of the junction is ramped at a constant sweep rate $dI/dt=17.5 \mu A/s$ and at least $10^4$ switching events have been recorded using a standard technique. These measurements, collected over a wide range of temperatures are reported in Fig.\ref{qpd4}a. A progressive broadening of the histograms occurs when  lowering the temperature, which is a distinctive feature of the phase diffusion regime\cite{kivioja2005,mannik2005,krasnov2005,fenton2008,Bae2009,luigi}. For temperatures below about 135mK, the histograms overlap.

The temperature dependence of the width $\sigma$ of the SCD curves is shown in Fig. \ref{qpd4}b.
As a test of fidelity for our fabrication process and our experimental setup we report data for two different samples with interface orientations of $75^o$ for sample A and $50^o$ for sample B. The junctions are 2.5 $\mu m$ and  2.0 $\mu m$ wide respectively, and the film thicknesses are 100 nm and 250 nm. We have selected interface orientations with robust overlap of the d-wave lobes on both sides of the junctions\cite{rop} and $E_J$ of the order of 3 meV. For both samples above a temperature $T^*$ the data display a decrease in the width of the distribution in agreement with a process of multiple escapes and retrapping typical of the diffusive motion. The high temperature region ($135 mK<T< 2 K$) allows a reliable estimation of the damping parameter. Simulations for  $Q= 1.30$ are reported as a red line in Fig. \ref{qpd4}b resulting from the integration of the Langevin equation (\ref{lang}) with a Bulirsh-Stoer integrator using a noise affected Runge-Kutta Monte Carlo algorithm\cite{luigi}.

In Fig. \ref{qpd4}c simulated thermal behavior of $\sigma$ is reported for different values of the Q damping parameter ranging from 1.2 to 5. For each of these curves $T^*$ approximately indicates the transition temperature from thermal activation to the diffusive regime. Q tunes $T^*$ as shown in the inset of Fig. \ref{qpd4}c and modifies the slope of the $\sigma(T)$ fall-off at higher temperatures. The capability to numerically reproduce this region makes it possible to estimate Q with high precision. In our case $Q= 1.30 \pm 0.05$ closely fits the data and determines  a $T^*$ value not larger than 100 mK. The section below $T^*$ faithfully reproduces the expected $T^{2/3}$ dependence for a thermally activated regime \cite{Jackel} (solid line) as an additional test of consistency. In Fig. \ref{qpd4}c the MQT section is missing. It would attach below $T_x$ to each of the curves with its characteristic saturation in $\sigma$, as shown in Fig. \ref{qpd4}b in fitting experimental data.

In facts below 135 mK $\sigma$ is almost independent of temperature and consistent with an escape dominated by quantum tunneling. Thus $135 \pm 10$ mK represents the cross-over temperature $T_x$  providing, once known the Q value from the phase diffusion fitting, first $C= 64$ fF and the plasma frequency  $\omega_p \simeq 38$ GHz. This value of the capacitance $C$ is also in agreement with the value extrapolated using the estimated specific capacitance $C_s \simeq 2 \cdot 10^{-5}$ $F cm^{-2}$  for YBaCuO BP JJs on LSAT substrates \cite{daniela,rop}. As a consequence $E_J \simeq 3 meV$ results to be much larger than $E_C \simeq 3 \mu eV$ ($E_J /E_C \simeq 1000$) and the system is in conditions nominally far from those which are considered to be favorable for the observation of quantum phase diffusion \cite{iansiti}.

In analogy to what commonly done to prove MQT in underdamped junctions\cite{devoret1985}, we use the magnetic field to tune in situ the junction parameters and $T_x$ to unambiguously prove MQT as source of the saturation of $\sigma$ below $T_x$.
In the inset of Fig. \ref{qpd4}b we report the temperature dependence of $\sigma$ measured for sample B  at two different magnetic fields of 0G and 12G respectively. H=12G lowers the critical current $I_{co}$ reducing at the same time the quantum crossover temperature. Relevant device parameters are summarized in Table \ref{table}.

\begin{table}
\caption{\label{table} Device parameters.}
\begin{ruledtabular}
\begin{tabular}{ccccccc}
\textbf{Sample}  & H (G)  &   $I_{co}$ ($\mu A$)  & R ($\Omega$) & C (fF)   &     $Q$  & $T_x$ (mK)  \\ \hline \\
A &  0 & 1.20 & 84  & 64  & $1.30 $ & 135   \\
B & 0  &  1.79 &  64  & 74 & $1.28 $ & 144  \\
B  & 12 &  1.42 & 64  & 74 & $1.14 $ & 122 \\

\end{tabular}
\end{ruledtabular}
\end{table}

\begin{figure}[htbp]
\begin{center}
\includegraphics[width=3.0in]{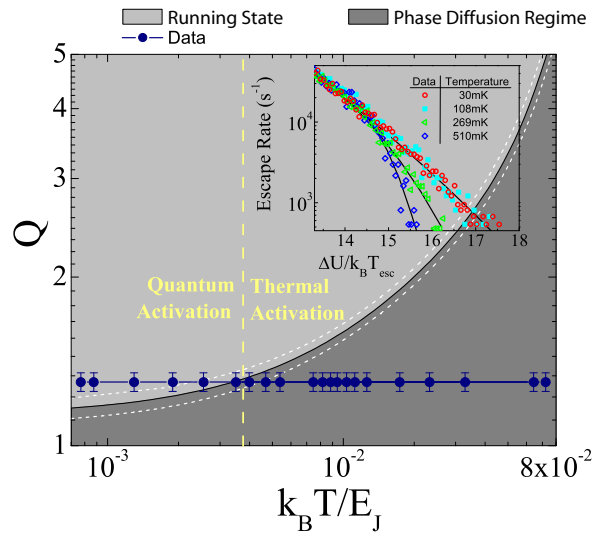}
\caption{$(Q, k_BT/E_J$) parameter space, showing the various activation regimes.
The transition curve between the PD regime and the running state has been extrapolated through numerical simulations, the sideband curves mark the uncertainty in our calculation and are due to the temperature step size. The crossover temperature between the various regimes scales with $E_J$. The data points refer to the Q values of our sample and show the direct transition from quantum activated running state to the PD regime. In the inset we show the experimental escape rates (symbols) as function of barrier height to escape energy ratio along with the theoretical fits at different T. The Q values used for the fits are the same shown in the phase diagram.}
\label{QPD2}
\end{center}
\end{figure}

In Fig. \ref{QPD2} we report a $(Q, k_BT/E_J)$ phase diagram, which summarizes the various activation regimes\cite{kivioja2005}. The transition curve between the phase diffusion regime and the running state following thermal (experimentally observed in \cite{kivioja2005,mannik2005,krasnov2005,luigi}) or quantum (experimentally observed in this work) activation has been determined numerically by varying the damping factor Q as function of the ratio between the thermal energy and the Josephson energy. The filled circles are experimental data obtained by fitting the escape rates $\Gamma$ as a function of the ratio between the barrier height and the escape energy, $u=\Delta U/k_B T_{esc}$ \cite{Note1} (shown in the inset of Fig. \ref{QPD2}). The obtained values fall within the region of the diagram that displays a direct transition from phase diffusion to quantum activation. The escape rates have been calculated from the switching distributions using a standard procedure\cite{fulton1974}. In the quantum activated regime the switching distributions are asymmetric and skewed to the left, and $\Gamma$ values all fall onto the same line, as it is the case for the reported data from T=30mK to 108mK. Retrapping processes cause a progressive symmetrization of the switching distribution which translates into a deviation of the experimental escape rates from the ideal exponential behavior\cite{fenton2008,luigi}.

In summary, by exploring a new region of the ($Q,k_BT / E_J$) phase diagram we have demonstrated a direct transition from quantum activation to diffusive Brownian motion in GB YBaCuO JJs. This experiment sets another milestone in the study of the influence of dissipation on the switching statistics of JJs\cite{chen,devoret1985,Ovch,vion,kirtley,castellano}demonstrating novel balancing between MQT, thermally activated and retrapping  escapes, thus paving the way to the observation of fully-quantum phase diffusion, and is of particular relevance to understand interaction of a quantum system with the environment\cite{Leggett}. The combined experimental and numerical investigation here presented has the potential to offer a new tool to study the interplay between coherence phenomena and dissipation down to the quantum regime in a wide variety of systems.

\begin{acknowledgments}
We thank T. Bauch and A. Ustinov for stimulating discussions. Special gratitude to J. Clarke and A.J. Leggett for inspiring conversations and AJL also for a careful reading of the manuscript. We acknowledge financial support by EC MIDAS and by a Marie Curie Grant n. 248933 "hybMQC". We also acknowledge support by ESF FoNE programme and by ISCRA the Italian SuperComputing Resource Allocation through grant IscrB\_NDJJBS 2011. We dedicate the work here presented to the memory of Prof. Antonio Barone who recently passed away. Antonio has been a fantastic friend, a great teacher, and a continuous source of inspiration for all of us. We have much profitted from conversations with him on the topics of this research.
\end{acknowledgments}

\end{document}